\documentclass[iop]{emulateapj}
\pdfoutput=1
\usepackage{graphicx}

\newcommand{\eg}{e.g.,\ }

\newcommand{\etal}{et~al.\ }
\newcommand{\ltsima}{$\; \buildrel < \over \sim \;$}
\newcommand{\simlt}{\lower.5ex\hbox{\ltsima}}
\newcommand{\gtsima}{$\; \buildrel > \over \sim \;$}
\newcommand{\simgt}{\lower.5ex\hbox{\gtsima}}

\newcommand{\magsec}{mag arcsec$^{-2}$}

\def\muv{$\mu_{\mbox{v}}$}

\def\bmv{B$-$V}

\def\metgrad{d[Fe/H]/dlog(r)}
\def\magdex{mag dex$^{-1}$}
\def\dexdex{dex dex$^{-1}$}
\def\re{$r_{\rm e}$}

\begin{document}

\title{Stellar Populations in the Outer Halo of the Massive Elliptical M49}

\author{J. Christopher Mihos,\altaffilmark{1} 
	Paul Harding,\altaffilmark{1}
	Craig S. Rudick,\altaffilmark{2} and
        John J. Feldmeier\altaffilmark{3}}
\email{mihos@case.edu, paul.harding@case.edu,
           craig.rudick@phys.ethz.ch, jjfeldmeier@ysu.edu}

\altaffiltext{1}{Department of Astronomy, Case Western Reserve University,
10900 Euclid Ave, Cleveland, OH 44106}

\altaffiltext{2}{Institute for Astronomy, ETH Zurich, CH-8093, Zurich, Switzerland}

\altaffiltext{3}{Department of Physics and Astronomy, Youngstown
State University, Youngstown, OH 44555, USA}

\begin{abstract}

We use deep surface photometry of the giant elliptical M49 (NGC 4472),
obtained as part of our survey for diffuse light in the Virgo Cluster,
to study the stellar populations in its outer halo. Our data trace M49's
stellar halo out to $\sim$ 100 kpc (7\re), where we find that the
shallow color gradient seen in the inner regions becomes dramatically
steeper. The outer regions of the galaxy are quite blue (\bmv$\sim$0.7);
if this is purely a metallicity effect, it argues for extremely metal
poor stellar populations with [Fe/H]$<$--1. We also find that the
extended accretion shells around M49 are
distinctly redder than the galaxy's surrounding halo, suggesting that we are
likely witnessing the buildup of both the stellar mass and metallicity
in M49's outer halo due to late time accretion. While such growth of
galaxy halos is predicted by models of hierarchical accretion, this
growth is thought to be driven by more massive accretion events which
have correspondingly higher mean metallicity than inferred for M49's
halo. Thus the extremely metal-poor nature of M49's extended halo
provides some tension against current models for elliptical galaxy
formation.

\end{abstract}


\section{Introduction}

Mounting evidence suggests that elliptical galaxies grow through a
two-phase process. The rapid assembly of gas-rich clumps at high
redshift likely results in a highly dissipative collapse process not too
dissimilar to ``monolithic collapse" models for elliptical galaxy
formation. This leads to the formation of dense, compact spheroidal
galaxies (Khochfar \& Silk 2006), similar to objects seen in recent high
redshift galaxy surveys (\eg Zirm \etal 2007; van Dokkum \etal 2008).
These systems are significantly more compact than local ellipticals, and
need to grow in size to become the ellipticals we see today. This argues
for a second assembly phase involving later accretion of low mass
objects with higher specific angular momentum in order to build up the
outer envelopes of the growing ellipticals (\eg Naab \etal 2007; Oser
\etal 2012).

Under this hierarchical model, the outskirts of nearby elliptical
galaxies should harbor a variety of signatures of this formation
process. Dissipative collapse should lead to strong metallicity
gradients (Carlberg 1984; Arimoto \& Yoshii 1987), but mergers
effectively mix the inner galaxy and largely wash out the gradients
(White 1980; Mihos \& Hernquist 1994). However, since this mixing is not
perfect, the outer halos should still retain a truly metal-poor
population of stars. Subsequent accretion of low mass galaxies would
then deposit stars with a broad range of metallicities into the outer
halo, building up the halo's surface brightness and mean metallicity.
Unlike the rapid mixing in the inner regions, the low density and larger
dynamical timescales in galaxy outskirts mean that the morphological
signatures of accretion -- shells and streams of stars -- survive
longer at large radius. Indeed, a variety of deep imaging studies show
that many ellipticals have extended tidal debris from late accretion
events (\eg Malin \& Carter 1983, Tal \etal 2009, Janowiecki \etal
2010).

While the structure of elliptical galaxy halos can be probed by deep
imaging, their extremely faint surface brightness makes detailed studies
of their stellar populations difficult. For sufficiently nearby
ellipticals, the discrete stellar populations can be imaged using the
{\it Hubble Space Telescope} (HST), giving strong constraints on the age
and metallicity of the stars (\eg Rejkuba \etal 2005, 2011; Harris \etal
2007). Stellar populations in more distant ellipticals must be studied
using their integrated colors or spectra. These studies generally show
that ellipticals have a gradual radial decline in the mean metallicity
over the inner few effective radii (\re), with 
gradients of \metgrad $\sim$ --0.1 to --0.3 
(\eg Peletier \etal 1990,
Kobayashi \& Arimoto 1999). However, most studies generally do not probe
the outer halo, where the stellar populations may have quite sub-solar
metallicities. Moreover, as the most massive ellipticals tend to live
near the center of galaxy clusters, studies of their outer halos are
complicated by contamination from the more extended intracluster
starlight (ICL) surrounding them. As such, the chemical abundance
constraints contained in the outer halos of giant ellipticals have
remained largely untapped. In this Letter, we present deep imaging of
the massive elliptical M49 ($\log M_*=11.9$, C\^ot\'e \etal 2003), which
lives in the outskirts of the Virgo cluster where ICL contamination is
minimized, to study the stellar populations in its outer halo.\footnote
{In this paper we adopt a distance of 16 Mpc (Mei \etal 2007) and effective radius of
194\arcsec (Kormendy \etal 2009) for M49.}

\section{Observations}

We observed M49 as part of our deep imaging survey of diffuse light in
the Virgo Cluster using Case Western Reserve University's 24/36" Burrell
Schmidt telescope. We give a brief description of the dataset here; more
details can be found in our earlier papers (Mihos \etal 2005; Janowiecki
\etal 2010; Rudick \etal 2010). Fields around M49 were imaged in Spring
2006 and 2007 in Washington M, and again in Spring 2011 in a custom B
filter; these filters are similar to standard Johnson B and V, but
$\sim$ 300\AA\ bluer. Each dataset consists of a large number of
overlapping images with individual exposure times of 1200s in B and 900s
in M, yielding sky levels of $\sim$ 800 and 1300 ADU respectively. The
images were flatfielded using a night sky flat constructed from 50--100
offset blank sky pointings. After subtracting the extended wings of bright
stars  and masking discrete sources on each image (stars, background galaxies, 
and the bright inner regions of large galaxies), sky subtraction was
achieved by subtracting planes fit to the unmasked background sky.

The photometric transformation to Johnson B and V magnitudes was derived
using Landolt UBVRI standards to measure the filter color terms, and the 
$\sim$ 200 stars on each frame with accurate SDSS photometry to derive
the photometric zeropoints (employing the Lupton 2005 $ugriz$ to Johnson 
transformation). The solutions showed a residual scatter of
0.02--0.03 mag in the photometry of individual stars and 0.01 mag
scatter in the frame-to-frame zeropoints. The individual images were
then scaled to a common zeropoint and median combined into final B and V
mosaics. In these mosaics, there are typically 30--50 and 40--90 images
contributing to any given pixel in the B and V mosaics, respectively.

At low surface brightness, accurate background subtraction and noise
estimation is critical to measurements of surface brightness and color.
To estimate and correct for uncertainty in the residual background
around M49 we identify 50 circular background apertures, 1.5\arcmin\ in
radius, located 30-60\arcmin\ from M49's center (beyond the area shown
in Figure \ref{images}). In these regions, the
average residual background is +0.4 ADU in B and $-$1.0 ADU in V, with
1$\sigma$ per-pixel noise of 0.75 and 1.0 ADU in B and V, respectively.
This gives us 2.5$\sigma$ limiting surface brightness of $\mu_{\rm
B,lim}=28.7$ and $\mu_{\rm V,lim}=28.3$, and we correct the photometric
data for Galactic extinction using values of $A_V=0.061$ and $A_B=0.081$
(Schlafly \& Finkbeiner 2011).

Finally, to increase signal-to-noise at low surface brightness the B and
V mosaics were masked and spatially rebinned to 13\arcsec\ resolution.
In this process, discrete objects on the mosaic were again masked, after
which the mosaics were rebinned into 9x9 pixel (13\arcsec x13\arcsec)
bins, calculating the median intensity of the unmasked pixels in each
bin. In the analysis that follows, the full resolution mosaic are used
to measure the photometric properties of M49's inner high surface
brightness regions, while the rebinned mosaics are used at larger radius
and lower surface brightness.

\section{Results}

\begin{figure}[]
\centerline{\includegraphics[width=3.5in]{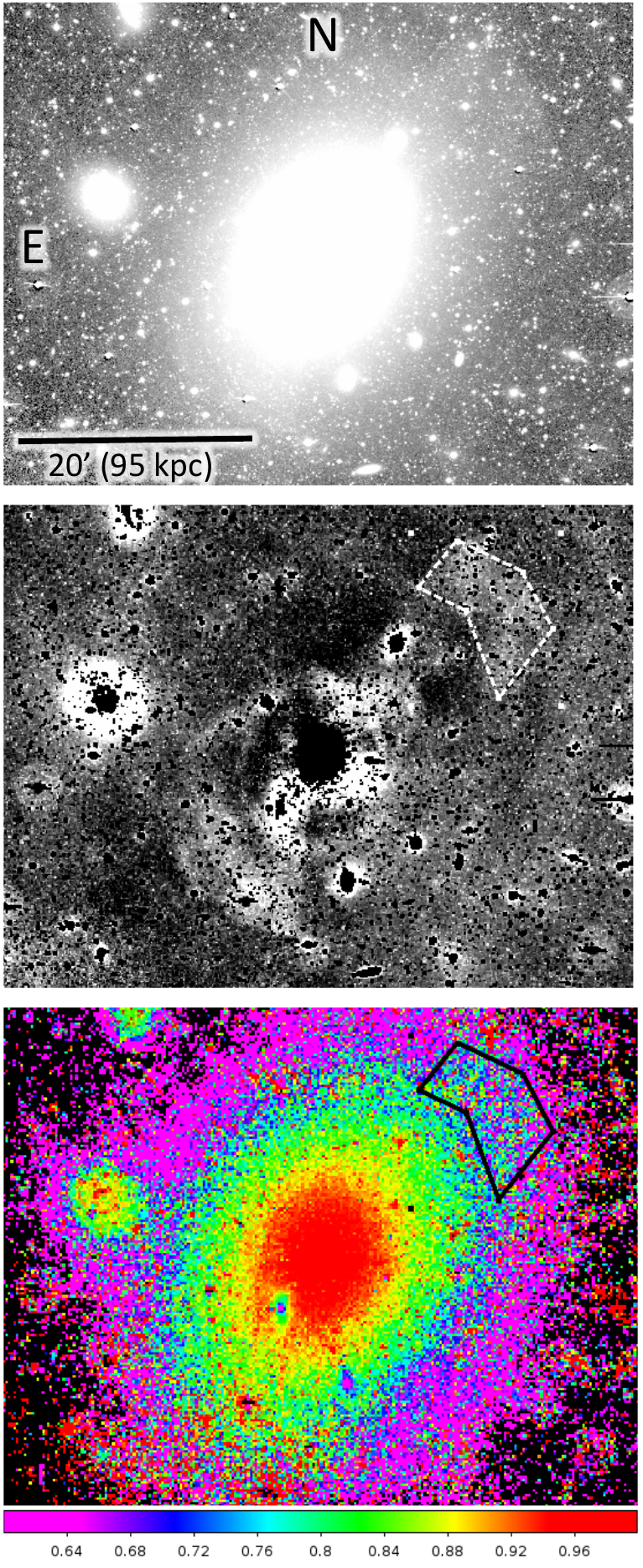}}
\caption{Top: V band image of M49. Center: Binned residual image after isophotal subtraction (from
Janowiecki \etal 2010) showing the tidal debris.
Bottom: \bmv\ color map, binned to 13\arcsec\ resolution. The NW Shell is marked.}
\label{images}
\end{figure}

Figure \ref{images} shows our V image of M49 (top) along with
a residual image (middle) adapted from Janowiecki \etal (2010),
constructed by subtracting an elliptical model of M49's light profile. 
A number of
accretion shells and streams can be seen surrounding M49; these are
discussed in more detail in Janowiecki \etal (2010) and have
been subsequently confirmed by Arrigoni Battaia \etal (2012). The bottom panel
of Figure \ref{images} shows our new \bmv\ colormap of M49, binned to
13\arcsec x13\arcsec\ resolution. In this image, we have ``unmasked'' the
high surface brightness regions of the galaxies to give visual
continuity to the image and show color across all surface brightnesses.

M49's color gradient can be clearly seen in the image, with colors
ranging from \bmv $\sim 1$ at 15\arcsec\ to \bmv$<0.7$ in the
outskirts ($r>1000$\arcsec).  Projected on the face of M49 are two 
blue star-forming companions, VCC1249 5\arcmin\ to the
southeast and NGC 4470 10\arcmin\ to the south. To the
northwest of M49, the extended debris shell seen in the residual map
can also be seen in the colormap; the shell region is
redder than the rest of M49's halo at similar radius ($r_{sma} =$
20\arcmin\ or 90 kpc). In contrast, the shells to the southeast do not show any
noticeable features in the colormap. This is not surprising
as these shells are at smaller radius ($r_{sma} =$ 12.5\arcmin\ or 60
kpc), where M49's halo is both brighter and redder. The southeastern
shells contribute less light to the color signal here and, if they are
similar in color to the northwest shells, would be similar in color to
M49's halo at that radius.

To extract the quantitative color profile of M49, we first mask M49's
companion galaxies out to their \muv\ $\sim$ 28.5 \magsec\ isophote as
measured in the residual image in Figure \ref{images}. We then calculate
the median surface brightness and color of all unmasked pixels in
elliptical annuli of constant ellipticity $\epsilon=0.28$ and position
angle PA$=-31$ (Kormendy \etal 2009; Janowiecki \etal 2010). Measured
this way, the derived surface brightness and color are areal-weighted rather than 
luminosity-weighted, and should be more indicative
of the diffuse halo, unbiased by contamination from any discrete
unmasked objects that fall within the annuli. 
The quartile errorbars on the outer points reflect the
effects of background subtraction uncertainty, and are calculated by
bootstrap sampling the background apertures and recalculating the
profiles for each background estimate.

\begin{figure*}[]
\centerline{\includegraphics[width=7.0truein]{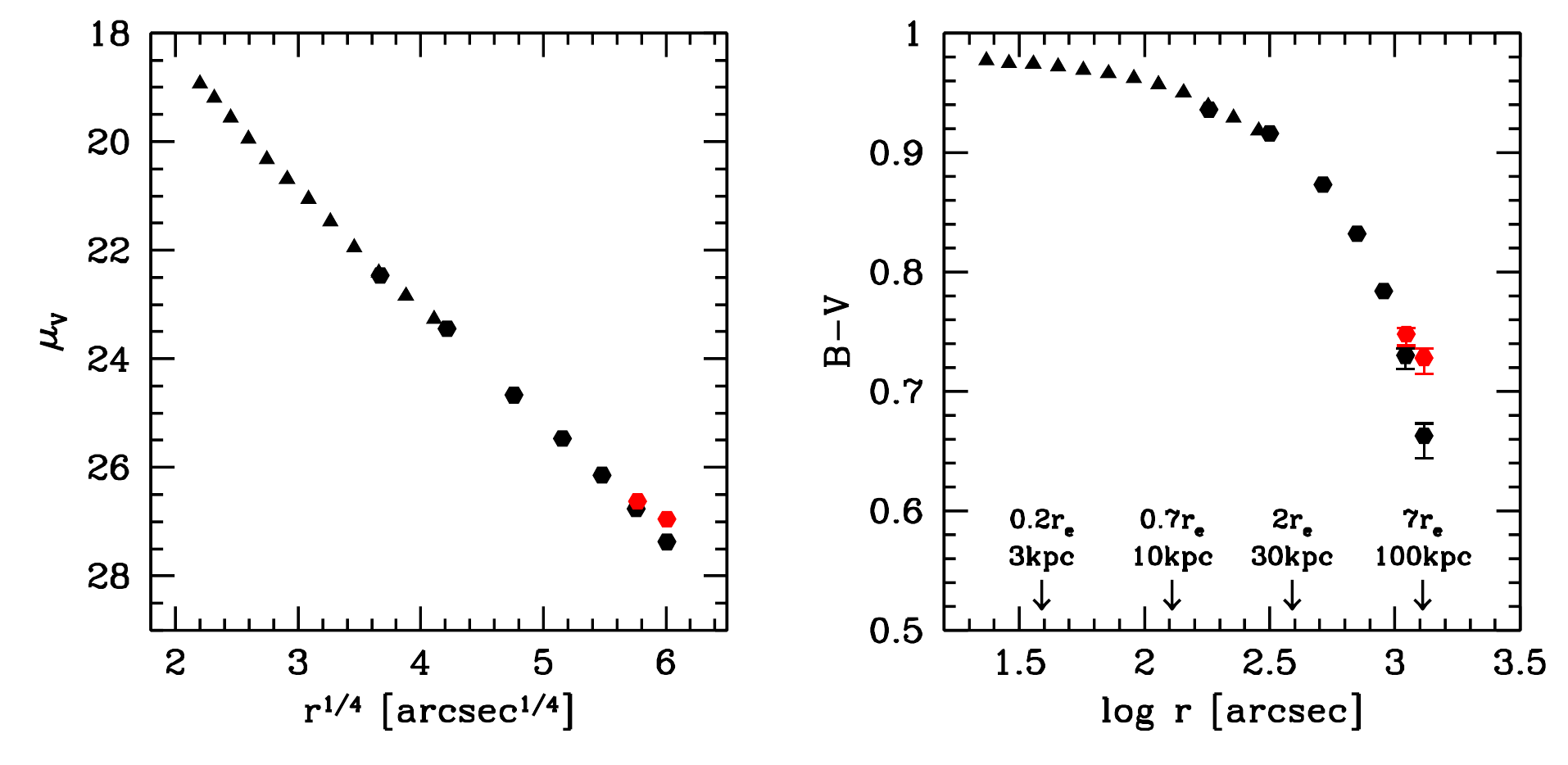}}
\caption{Left: Surface brightness profile of M49. Right: Color
profile. Black points represent azimuthally-average profiles;
triangles are measurements made from the mosaic at native 1.45\arcsec\
resolution, while hexagons show the profiles computed after re-binning to 13\arcsec\ resolution. Errorbars
represent the effect of background uncertainty on the profiles, and are
only larger than the point size in the outermost points in the color
profile. In both panels, red points denote measurements made along an
angular wedge containing the NW Shell.}
\label{colorprof}
\end{figure*}

M49's color profile (Figure \ref{colorprof}) shows a very shallow 
gradient in the inner regions which becomes much steeper at large
radius. Inside of r=100\arcsec\ (8
kpc), the logarithmic color gradient is $\Delta (B-V) \equiv d(B-V)/d\log r = -0.03$ \magdex, but the gradient steepens
continuously with radius, reaching values of --0.3 \magdex\ out at
r=800\arcsec\ (64 kpc). At the outermost point we measure, the color has
dropped to an extremely blue \bmv=0.66$\pm$0.02.

Inside an effective radius, our photometry compares well to previously
published data for M49, all of which show a systematic bluing of the
colors with increasing radius. Our \bmv\ colors and gradient match those
measured by Idiart \etal (2003), and, assuming old stellar populations,
are commensurate with gradients measured in other optical colors (Bender
\& Mollenhoff 1987, Peletier \etal 1990, Kim \etal 2000). At larger
radius, however, the situation becomes more complicated. Cohen (1986)
measure continual bluing in $g-i$ out to $\sim$ 350\arcsec, but a
reddening in $g-r$, while Kim \etal (2000) show a gradual reddening from
200\arcsec--500\arcsec\ in C--T$_1$. At these radii, M49's surface
brightness is well below sky, making the derived color
gradients very sensitive to accurate sky subtraction. In our
data, however, the rapid change in slope becomes clearly noticeable at 30 kpc
(2\re), where the sky uncertainty is only $\sim$ 1\% of the measured
surface brightness. Even in our outermost radial bin, the surface
brightness is a magnitude above the per-pixel sky noise, and the
errorbars on those points reflect the effects of the global sky
uncertainty. Clearly, errors in sky subtraction are not the cause of the
rapid bluing we see at large radius in M49.

\section{Discussion}

\begin{figure}[]
\centerline{\includegraphics{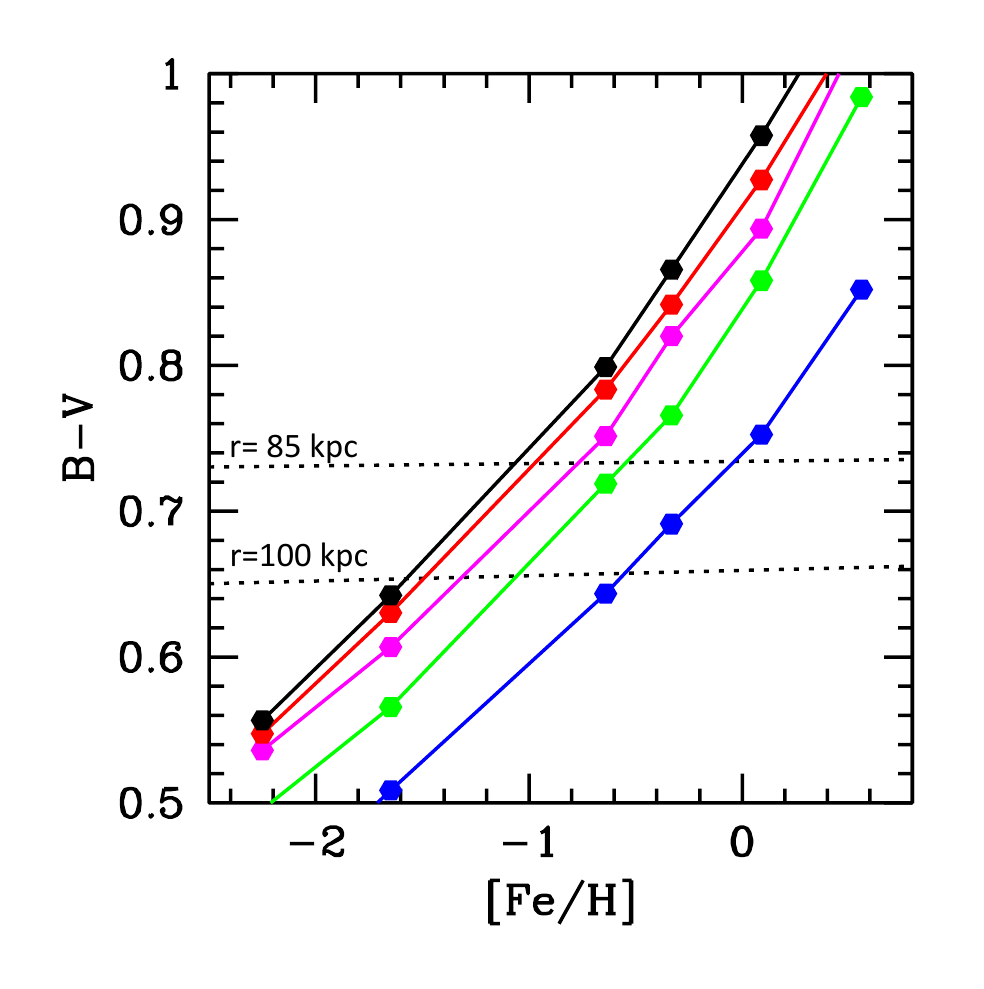}}
\caption{Simple stellar population models from Bruzual \& Charlot
(2003). From top to bottom, the curves show models with ages 10,
8, 6, 4, and 2 Gyr. The horizontal dashed lines show the color of the two outermost 
points in our profile. }
\label{SSPmodels}
\end{figure}

Color trends in elliptical galaxies -- both radial gradients within
galaxies and trends between color and luminosity in galaxy populations
-- are well-established to be driven largely by metallicity effects. Figure \ref{SSPmodels} shows the relationship
between \bmv\ color and metallicity for single stellar population (SSP)
models from Bruzual \& Charlot (2003), using a Chabrier IMF and Padova
isochrones. For [Fe/H]$>-1$ and populations older than a few Gyr, our
inner color gradient translates to a metallicity gradient of
$\Delta$[Fe/H] = --0.15 \dexdex, similar to the metallicity gradient
derived by Peletier \etal (1990) using B--R colors, but somewhat more
shallow than those derived spectroscopically, which suggest metallicity
gradients of --0.2 to --0.3 \dexdex\ (Kobayashi \& Arimoto 1999). The
gradient steepens in the outskirts; if interpreted solely as a
metallicity effect, by the inferred gradient reaches $\Delta$[Fe/H] =
--2 to --3 \dexdex\ by 1000\arcsec. 
The rapid steepening of the color profile begins at approximately the same
radius where the luminosity profile begins to show excess light above
a pure $r^{1/4}$ law, perhaps suggesting that instead of a simple
steepening of the metallicity gradient, we may be seeing a more discrete transition
from a metal-rich component to a metal-poor component (\eg Harris \etal
2007). 
For old populations, metallicities in the range [Fe/H]=--1 to
--1.5 are needed to explain the extremely blue colors of the outer
isophotes, metallicities similar to those of M49's metal-poor globular
cluster system ([Fe/H]$\sim$--1.3; Geisler \etal 1996, Cohen \etal
2003).

In principle, the steep gradient could also be explained by
systematically younger population ages at large radius. If we
extrapolate the shallow inner metallicity gradient outwards, we
would expect metallicities in the outer halo of [Fe/H] $\sim$ --0.15 to
--0.3. At these metallicities, the only way to match the outer isophotal
colors is using stellar populations with SSP-equivalent ages of
$\sim$ 2 Gyr (Figure \ref{SSPmodels}). This is unrealistically small;
while the recent accretion of a late-type galaxy would lead to younger
inferred ages, it would have to be an extremely massive and recent event
to so dramatically affect the luminosity-weighted age. Given that the
accretion features we see in M49 amount to only $\sim$ 1\% of the
galaxy's luminosity (Janowiecki \etal 2010) and are morphologically akin
to those expected from a minor merger (Hernquist \& Quinn 1989), such a
scenario is unlikely. Adopting a moderately lower halo metallicity of
[Fe/H] $\sim$ --0.7, raises the inferred population ages to $\sim$ 3--6
Gyr, still quite young for a SSP-equivalent luminosity-weighted age.
Halo metallicities of [Fe/H]$\leq-1$ are thus the most likely
explanation for the blue colors of M49's outer halo.

How do these stellar population constraints compare to those in the
outskirts of ellipticals more generally? A systematic comparison is
difficult, as few
studies have probed elliptical halos beyond a few \re. What information does exist shows
a wide range of properties: in some systems the inner metallicity
gradients continue smoothly out to 2--4 \re\ (Weijmans \etal 2009; Greene
\etal 2012), while others show gradients which either steepen or flatten
at large radius (Baes \etal 2007; Foster \etal 2009). Inferred halo
metallicities span the range $-1.5<$ [Fe/H] $< 0$ (Baes \etal 2007,
Foster \etal 2009, Greene \etal 2012). while most age estimates tend to
yield older ages in the outer halo (8-12 Gyr; Baes \etal 2007; Greene
\etal 2012). In cases where younger ages are inferred, there is evidence
for a recent accretion event in the galaxy's halo (\eg NGC 3348; Baes
\etal 2007). While our data suggest that M49's outskirts are at the
extreme end of the inferred age/metallicity ranges in these other
studies, we stress the very large radii being probed; restricting our 
analysis to the inner few \re\ would show only a modest population
gradient similar to those previously determined.

For a few nearby ellipticals, additional information comes from HST
studies of resolved stellar populations in their outer halos. NGC 5128
(Centaurus A) shows a mean metallicity of [Fe/H]=--0.5 at $r$=40 kpc
(7\re, Rejkuba \etal 2005), while NGC 3379 shows an extremely broad
metallicity distribution at r=33 kpc (12\re) with a mean metallicity of
[Fe/H] = --0.7 (Harris \etal 2007). These metallicities are comparable
to what we infer (assuming old stellar populations) for M49 at similar
{\it physical} radius ($r$=30--40 kpc), but higher when compared at
similar scaled radius ($r/$\re$\sim$10). NGC 5128 also shows evidence for a second
population of younger stars (ages $\sim$ 2--4 Gyr) in its stellar halo
(Rejkuba \etal 2011), likely related to the galaxy's status as a post
accretion system. However, these young stars comprise at most 10\% of
the mass of the halo, and the inferred integrated \bmv\ color of NGC
5128 would still be redder than what we observe in the outskirts of M49.

Theoretical expectations for the stellar populations in elliptical halos
are similarly diverse. While an initial dissipative collapse 
should yield strong metallicity gradients (Larson 1974; Carlberg 1984;
Arimoto \& Yoshii 1987), subsequent merging effectively mixes the
inner regions (White 1980; Mihos \& Hernquist 1994;
Kobayashi 2004), leaving behind the weak gradients observed inside a few
\re. However, chemodynamic simulations of elliptical galaxy formation
(Kobayashi 2004) show that the outer metallicity gradients can remain
quite steep, yielding metallicities of [Fe/H] $<-1$ at large radius. Of
course, additional accretion of satellite systems can deposit stars of
varying age and metallicity to the outer halo, leading to wider
diversity in the halo stellar populations.

Indeed, our photometry directly shows the effect of satellite accretion
on the halo populations in M49. The NW Shell is $\sim$ 0.07 mags redder
than the surrounding halo; since this region consists of light from both
the shell and from M49's (bluer) halo, the intrinsic color of the Shell
must be even redder, with \bmv $\sim$ 0.85. The fact that the shells are
redder than the surrounding halo can be understood in terms of a
disruptive accretion event. Because of the mass-metallicity
relationship, the mean color of the accreted satellite will be bluer
than that of M49 as a whole, but can be redder than M49's outer halo.
For example, van Zee \etal (2004) show that the
population of dE's in Virgo has a mean B-V color of 0.77 at $m_B = 15$
(roughly 1\% the luminosity of M49); if a system like these were
accreted by M49, it would leave a shell system much like what is shown
in Figure \ref{images}.

In the picture painted here, M49's metal-poor halo is a relic of the
rapid early assembly of the galaxy, while ongoing accretion (as seen in
the tidal shells) continues to build the halo mass and metallicity over
time. However, this scenario is not without problems. With a velocity
dispersion of 280 km s$^{-1}$ and effective radius of 15 kpc, M49 sits
squarely on the mass-size relationship for galaxies in the local
universe (van~der~Wel \etal 2008). In other words, M49 has {\it already}
built its halo to modern-day standards, but has done so while retaining
its metal-poor nature. Simulations suggest that the dominant mode of
halo growth is via minor mergers with typical mass ratios of 1:5 (Oser
\etal 2012); such mergers likely would have deposited stars with
significantly higher metallicity than that inferred for M49's halo. This
tension between a fully-developed halo and its low metallicity remains
unresolved.

It is interesting in this context to compare M49 to M87, the giant
elliptical at the heart of Virgo. While the two galaxies are comparable
in luminosity, M87 has a higher Sersic index ($n=11.8$, compared to
M49's $n=6.0$; Kormendy \etal 2009) and flatter color gradient (Rudick
\etal 2010) at large radius than M49. Where it sits, M87 is also subject
to a constant bombardment of satellite galaxies; this dynamically active
environment continually adds material to M87's outer halo, as well as
the ICL around M87 (Mihos \etal 2005), resulting in the more extended
envelope with relatively high mean metallicity. In contrast, M49 lies on
the outskirts of the Virgo Cluster, projected 1.2 Mpc from the cluster
center, and may be falling into Virgo for the first time (Irwin \&
Sarazin 1996; Janowiecki \etal 2010). As such, its halo may be more
indicative of ellipticals in field and group environments, and be less
``processed" than cluster ellipticals.

Stellar populations in the halos of ellipticals hold important
information on the processes shaping today's elliptical galaxies. 
If M49's extended but extremely metal-poor halo is a common
feature in massive ellipticals,  a revision of galaxy formation models may be necessary. Progress on
these issues requires a better census of the halo stellar populations in
ellipticals. Unfortunately, broadband colors are a very blunt tool for
constraining stellar populations, unsuited for disentangling age and
metallicity effects, while the very low halo surface brightness makes
more finely-tuned spectroscopic studies difficult. More promising is the
study of discrete stellar populations. At the distance of Virgo, deep
HST imaging can reach sufficiently far down the red giant branch to
provide constraints on the stellar populations, and has been used to
study the metallicity distribution in Virgo's intracluster light
(Ferguson \etal 1998; Durrell \etal 2002; Williams \etal 2007). A
similar study of the outer halos of M49 and other Virgo ellipticals, as
well as ellipticals in the nearby field, would provide a direct test of
the extremely metal-poor populations inferred from our deep imaging, and
open up a new avenue for the study of elliptical galaxies.

\acknowledgments

This work has been supported by the NSF through grants AST-0607526 and 
AST-1108964 to JCM and AST-0807873 to JJF. We thank Heather Morrison, Scott Trager, and Antonio
Pipino for many useful discussions.

{\it Facility:} \facility{CWRU:Schmidt}

\end{document}